\documentclass[twocolumn,letterpaper,prb]{revtex4}
\usepackage{amsmath,amsthm,amscd}
\usepackage{dcolumn}
\usepackage{graphicx}
\usepackage{epsfig}
\begin{document}
\title{Coherent control of lattice deformations in quantum
wires by optical self-trapping} \author{M. V. Katkov and
C. Piermarocchi} \affiliation{Department of Physics and Astronomy,
Michigan State University, East Lansing, Michigan 48824}

\begin{abstract}
We investigate theoretically a semiconductor quantum wire under the
effect of an intense off-resonant cw laser.  We show that in the
regime of strong exciton-phonon coupling the light-dressed ground
state of the wire reveals symmetry-breaking features, leading to local
lattice deformations.  Due to the off-resonant nature of the
excitation, the deformations are reversible and controllable by the
intensity and frequency of the laser. We calculate the light-induced
strain forces on the lattice in the case of an organic quantum wire.
\end{abstract}

\date{\today} \maketitle 

When an atomic system or a semiconductor is
irradiated by a pump laser in the transparency spectral region, it
responds to the field with its dynamic polarizability and gains a
polarization energy.~\cite{cohen92} In the case of semiconductors,
this dynamic Stark effect is well understood in terms of the creation
of virtual electron-hole pairs (excitons) across the band gap by pump
photons,~\cite{combescot88} and it explains the excitonic blue shift
experimentally observed.~\cite{mysyrowicz86} It has been recently
pointed out that the dynamic polarization can be used as a quantum
control tool in semiconductors. For instace, the virtual excitons
created by an off-resonant pump field interact with spins localized by
impurities or quantum dots and can create a local magnetic
field,~\cite{combescot04} induce spin-spin
coupling,~\cite{piermarocchi04,piermarocchi02} and paramagnetic to
ferromagnetic transitions.~\cite{fernandez04}

In this paper we will show that the dynamic optical polarization can
be used to induce strain forces and control lattice deformations in a
semiconductor quantum wire. If the virtual excitons created by the
pump field interact strongly with phonons in the lattice, a
self-trapping~\cite{toyo} of the optical polarization results. This
induces local forces on the lattice.  As in the case of the optical
spin control, there is no absorption of energy since the pump photons
are tuned in the transparency region: the effect is due to radiative
(stimulated) corrections to the ground state of the system. The
coherent nature of the effect makes it reversible and finely
controllable with lasers. In principle many different materials have
the exciton-phonon coupling strong enough for this optical
self-trapping effect.~\cite{song93} Here we focus on an organic one
dimensional system, polydiacetylene (PDA),~\cite{book} which has been
studied for its strong nonresonant optical nonlinearities and
exciton-phonon effects. Phonon-mediated optical nonlinearities have
been observed in this material.~\cite{greene90}

Recently-developed experimental techniques can detect light-induced
lattice displacement in molecules and semiconductor systems. Examples
include pump-probe electron diffraction~\cite{ihee01} and ultrafast
x-ray absorption spectroscopy.~\cite{bressler04} These techniques can
be extremely sensitive and measure laser-induced lattice dynamics
within picosecond and milli-{\AA}ngstr\"om
resolution.~\cite{rosepetruck99} PDA has been already identified as a
good system where the vibrational dynamics after optical excitation
could be observed,~\cite{tanaka03} and the possibility of addressing
optically a single polymer chain of this material has been recently
demonstrated.~\cite{guillet01} We are focusing here on the steady
state regime of the polymer driven by a cw or nanosecond laser. This
should be easier to address experimentally, yet it contains rich and
unexplored features related to light-matter interaction in strong
coupling. We remark that, in contrast to more common schemes based on
photothermal effects, the coherence of the electronic excitation is a
key element in this light-induced effect.~\cite{stoneham03}

The coherent many-body ground state of the wire in the presence of the
light field can be parametrized as a BCS-like wavefunction that
depends on variational parameters. By a functional minimization of the
ground state energy with respect to these variational parameters, we
obtain a non-homogeneous nonlinear equation that describes the
distribution of the optical polarization in the wire.  Unlike the case
of a single exciton-polaron,~\cite{W1} the total polarization is not
fixed but is determined by the intensity and frequency of the laser
field. The nonlinear equation is solved numerically to calculate the
distribution of the polarization and therefore the strain forces
acting on the wire. Recent esperiments on single PDA
chains~\cite{guillet01,DBG,lecuiller02} suggest that free excitonic
states have a Bohr radius of 1-2 nm, corresponding to 2-4 monomer
molecules, and a mass of $0.3 m_0$.  We use these parameters to
describe the wire with a tight-binding model. The model takes
explicitly into account the exciton-light coupling. The hamiltonian
can be written as
\begin{equation} 
H=H_{SSH}+H_{XL}~,
\label{hh}
\end{equation}
where the first term has a Su-Schrieffer-Heeger (SSH)~\cite{SSH} form   
\begin{eqnarray}
H_{SSH}&=&\sum_{n}\frac{p_n^2}{2M} + \sum_n
\frac{C}{2}(u_{n+1}-u_n)^2 \nonumber \\ 
& & -\sum_n
t_{n+1,n}(B^\dagger_{n+1}B_n +B^\dagger_n B_{n+1})~,
\label{eqssh}
\end{eqnarray} 
and  the exciton-light interaction is described by
\begin{equation}
H_{XL}=\sum_n
\frac{\Omega}{2}(B^\dagger_n+B_n) +\sum_n
(E_g+ 2 t_0-\hbar\omega_L)B^{\dagger}_n B_n~.
\end{equation}
$M$, $C$, $u_n$ and $p_n$ are the mass, elastic constant, total
displacement and momentum of the $n$-th site of the tight-binding
chain.  $B^+_n$, $B_n$ are operators of creation, annihilation of
excitons in a singlet spin state.  The hopping term is
$t_{n+1,n}=t_0-\gamma(u_{n+1}-u_n)$, where $\gamma$ is related
exciton-phonon deformation potential interaction: $D=2 \gamma a$. $a$
is the site separation in the tight-binding chain taken equal to the
excitonic Bohr radius (1.5 nm for PDA).  The value of $t_0$ is
determined by the exciton effective mass $m$ as $ t_0=\hbar^2/2m a^2$.
We treat the exciton-light coupling ($H_{XL}$) semiclassically, and
the rotating wave approximation is assumed. The shift $2t_0$ is
introduced in such a way to define $E_g$ as the $k=0$ exciton energy
at the bottom of the band, and $\omega_L$ is the frequency of the
laser. $\Omega= d\mathcal{E}_0$ is the Rabi energy due to the dipolar
coupling of excitons with a light field of amplitude
$\mathcal{E}_0$.  We remark that PDA has a large exciton binding
energy (of the order of 500 meV) and also shows vibronic resonances
spectrally localized at about 200 meV below the free exciton
peak.~\cite{lecuiller02} If the optical detuning $\delta=
E_g-\hbar\omega_L$ is small compared to these quantities, both free
e-h pairs and vibronic excitations can be neglected in the model.

In contrast to the polyacetylene case,~\cite{SSH} the ground state of
polydiacetylene is a non-degenerate state that we can express in the
form
\begin{equation}
|\Phi_0\rangle=\prod_{n} |0\rangle_{n}
\label{psi0}
\end{equation}
meaning that the wire has no excitons. However, in the presence
of light, the coherent light-dressed ground state will change, and we
write it using the ansatz~\cite{eastham01}
\begin{equation}
|\Phi_{\Omega}\rangle=\prod_{n}
 \left(\alpha_n|0\rangle_{n}+\beta_ne^{i\varphi_n}|1\rangle_{n}\right)~,
\label{var}\end{equation}
where $\alpha_n$, $\beta_n$, and $\varphi_n$ are variational
parameters, and $\alpha_n$ and $\beta_n$ are subject to the
single-occupancy constraint
$|\alpha_n|^2+|\beta_n|^2=1$. $|0\rangle_n$ and $|1\rangle_n$ denote
the number of excitons in the site $n$.  Due to the hopping, the
variational coefficients depend on the index $n$. We introduce the
phases $\varphi_n$ to make $\alpha_n$ and $\beta_n$ real.  The mimimum
of the total energy occurs when the phases are constant, making all
the two-level oscillators in the chain mutually
coherent.~\cite{katkov05} Using the wavefunction in Eq.(\ref{var}) and
the full Hamiltonian in Eq.~(\ref{hh}), we calculate the total
energy. By introducing the optical polarization function
\begin{equation}
\psi_n=-\langle B_n \rangle=2\alpha_n\beta_n~,
\end{equation} 
the energy can be
written as
\begin{eqnarray}
-\frac{1}{2}\left(\sum_n
 t_{n+1,n}\psi_{n+1}\psi_n+\Omega\sum_{n}\psi_n+\delta^\prime\sum_{n}\sqrt{1-\psi^2_n}\right),
\label{sign}
\end{eqnarray}
(terms that do not depend on $\psi_n$ have been dropped) where we have
defined $\delta^\prime=\delta +2t_0$. We also have set
$\varphi_n=\pi$, and assumed $\alpha_n\beta_n>0$ and
$|\alpha_n|>|\beta_n|$~.  By functional derivation with respect to
$\psi_n$ we obtain
\begin{equation}
-t_{n+1,n}\psi_{n+1}-t_{n,n-1}\psi_{n-1}-\Omega
 +\delta^\prime\frac{\psi_n}{\sqrt{1-\psi^2_n}}=0 \label{varofeq}~,
\end{equation} which in the continuum limit reads
\begin{equation}
-t_0 \left(2\psi_n+\frac{\partial^2 \psi_n}{\partial
 n^2}\right)+2\gamma\psi_n\frac{\partial u_n}{\partial n}-\Omega
 +\delta^\prime\frac{\psi_n}{\sqrt{1-\psi^2_n}}=0 \label{twoterms}.
\end{equation} 
In order to have a closed equation for $\psi_n$ we need to find
$\partial u_n/\partial n$, which depends self-consistently on the
distribution of the polarization along the chain in the steady
state. This can be calculated using the same approch of Ref.~\onlinecite{W1}
by writing the classical equations of motions for the wire
\begin{equation}
M\ddot{u}_n=-\frac{\partial \langle H \rangle }{\partial
u_n}=C(u_{n+1}-u_n)-C(u_n-u_{n-1})+F_n~.
\label{classicaleq}
\end{equation}
In the continuum limit the strain force on the site $n$, $F_n$, is
related to the gradient of the polarization density as
\begin{equation}
F_n=\frac{\gamma}{2} \frac{\partial|\psi_n|^2}{\partial n}~.
\label{forceeq}
\end{equation}
Since we are considering a stationary solution in the presence of the
light we set Eq.~(\ref{classicaleq}) to zero, which gives the local
displacement due to the strain force as $u_n=-C^{-1} \int \int
F_n\,d^2n$.  Then from Eq.~(\ref{forceeq}) we find
\begin{equation}
\frac{\partial u_n}{\partial n}=-\frac{\gamma}{2C}|\psi_n|^2 +a\Delta
\label{derofdisp}
\end{equation}
where $ \Delta$ is a dimensionless constant of integration that gives
finite-size effects. We choose $\Delta=0$, which implies that the
total lenght of the wire is not fixed. This choice does not affect the
results in the limit of a wire much longer than the extension of the
self-trapping region. The force constant $C$ can be expressed in terms
of the sound velocity $S$ as $C=S^2M/a^2$ where $S=2.5\times10^3$ m
s$^{-1}$ for PDA.~\cite{DBG} Finally, from Eqs.~(\ref{twoterms}) and
(\ref{derofdisp}) we get the equation for $\psi_n$
\begin{widetext}
\begin{equation}
 2t_0\left(\frac{1}{\sqrt{1-\psi^2_n}}-1\right) \psi_n -t_0
 \frac{\partial^2\psi_n}{\partial
 n^2}-\frac{D^2}{4MS^2}\psi_n^3-\Omega
 +\delta\frac{\psi_n}{\sqrt{1-\psi^2_n}}=0.\label{final}
\end{equation}
\end{widetext}
Notice that we are dealing with a system where the quantity $\psi_n$
is not normalized. The amount of total polarization density in the
wire, i.e. the number of virtual excitons, is determined by the
external field parameters $\Omega$ and $\delta$.  The first term in
Eq.~(\ref{final}) describes the blue shift of the exciton (Pauli
blocking).  The cubic term, related to the exciton-phonon deformation
potential interaction $D$, is responsible for the self-trapping of the
polarization. It corresponds to an attractive potential proportional
to the polarization density. The last term takes into account the
effect of the finite optical detuning of the laser field. In the
absence of light (i.e. for $\Omega=0$), the solution to
Eq.~(\ref{final}) is $\psi_n=0$ corresponding to the trivial solution
in Eq.~(\ref{psi0}).  In the absence of hopping between the sites
($t_0=0$, $D=0$), we have spatially separated excitations with no
constraints applied.~\cite{eastham01} The solution in this case
corresponds to the one of the dynamic Stark effect in
atoms~\cite{cohen92}
\begin{equation}
\psi_n=\frac{\Omega}{\sqrt{\delta^2+\Omega^2}}.
\label{atoms}
\end{equation}
At resonance ($\delta=0$) Eq.~(\ref{atoms}) gives the maximum
occupation $\psi_n=1$, corresponding to half exciton per site, and the
light-dressed ground state is $2^{-1/2}(|0\rangle-|1\rangle)$ on each
site. In the limit of small excitation $\psi_n \ll 1$,
Eq.~(\ref{final}) is similar to a one-dimensional nonlinear
Schr\"odinger equation with a cubic attractive nonlinearity. It is
known that a Bose-Einstein Condensate (BEC) with an attractive
nonlinearity breakes symmetry in 1D so that the ground state does not
correspond to a constant amplitude solution.~\cite{carr00} In our
case, these real symmetry-breaking solutions for the ground state
describe the optical self-trapping effect. In the small excitation
limit some analytical forms for the solution of Eq.~(\ref{final}) can
be written in terms of Elliptic functions using Cayley's
transformations.~\cite{katkov05} It is interesting to remark that the
Hamiltonian in Eq.~(\ref{hh}) can be mapped to an XY model for spins
coupled to phonons. The Rabi energy describes an external magnetic
field in the XY plane while the optical detuning corresponds to a
magnetic field in the Z direction.  An effective nonlinear interaction
between spins analogous to our cubic term was found in the case of a
classical Heisenberg spin chain coupled to phonons.~\cite{barma75}

In order to study on the same ground the inter-cell hopping and the
presence of electromagnetic field we solve numerically
Eq.~(\ref{final}). This allow us to calculate the forces on the
lattice for an arbitrary value of the external field, provided the
system remains coherent. To solve the equation numerically, we use the
steepest descent method of functional minimization.~\cite{SCP} This
method is efficient in solving numerically Gross-Pitaevskii equations
for BECs, which are in the same class of
Eq.~(\ref{final}).~\cite{dalfovo96} We start from an initial trial
function $\psi_n(0)$, then $\psi_n(\tau)$ is evaluated in terms of the
equation$ \frac{\partial \psi_n(\tau)}{\partial \tau}=-{\cal{H}}
\psi_n(\tau)$~, by choosing an arbitrary step $\triangle \tau$ in the
fictitious time variable $\tau$ and by iterating according to
$\psi_n(\tau+\triangle \tau)\approx \psi_n(\tau) -\triangle \tau
{\cal{H}} \psi_{n}(\tau)$.  The nonlinear operator ${\cal{H}}$ is
defined in such a way that ${\cal{H}} \psi_n(\tau)=0$ corresponds to
Eq.~(\ref{final}).  At convergence, the condition $\frac{\partial
\psi_n}{\partial \tau}=0$ gives the solution to Eq.~(\ref{final}).
\begin{figure}
\includegraphics{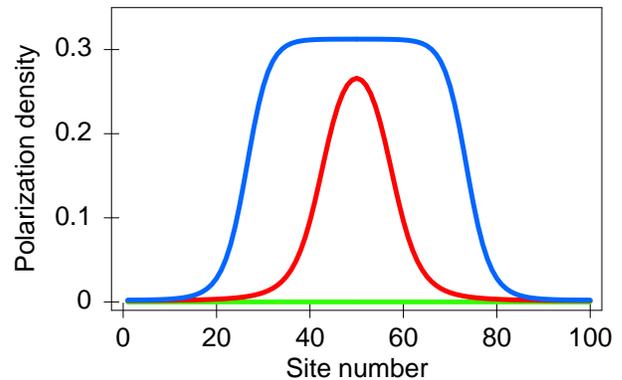}
\caption{\label{polar} (Color online) Polarization density
$|\psi_n|^2$ for $\delta/t_0=0.05$, $\Omega/t_0=10^{-4}$ (green line),
$\Omega/t_0=1.99 \times 10^{-3}$ (red line), and
$\Omega/t_0=2.06\times 10^{-3}$ (blue line)}
\end{figure}

\begin{figure}
\includegraphics{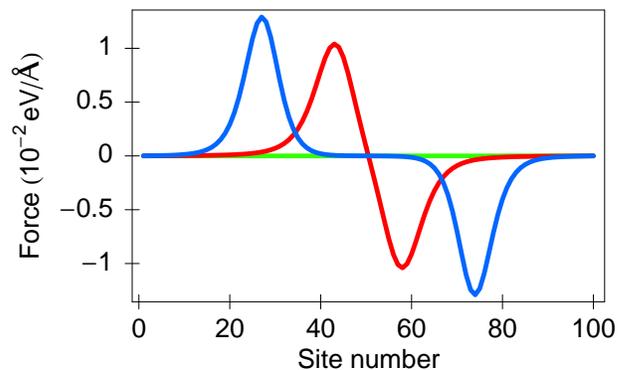}
\caption{\label{force} (Color online) Force as a function of the site
number for $\delta/t_0=0.05$, $\Omega/t_0=10^{-4}$ (green line),
$\Omega/t_0=1.99 \times 10^{-3}$ (red line), and
$\Omega/t_0=2.06\times 10^{-3}$ (blue line)}
\end{figure}

The results of the numerical solution for $\psi_n$ are shown in
Fig.~\ref{polar}. The linear density of the wire is $ 1.4 \times
10^{-14}$ g/cm.~\cite{W1}. The sum of the deformation potentials for
the conduction and valence band is $D=D_c+D_v=6.1$ eV.~\cite{DBG} The
dipole momentum of PDA is 10.1 e\AA,~\cite{Wei} and the energy of the
excitonic (singlet) state is 2.282 eV.~\cite{guillet01} The numerical
solution is obtained in a chain with 100 sites and periodic boundary
conditions are used.  The total lenght of the chain is  smaller
than the wavelength of the light and we can neglect the effect of a
finite photon wavevector.  The experimental value of the excitonic
linewidth at 10 K is 350 $\mu$eV,~\cite{guillet01} which is much
smaller than the values for the Rabi energy and detuning we use in the
calculations. We have also checked that the exciton-polaron binding
energy is always smaller than $\delta$. Fig.~\ref{polar} shows
$|\psi_n|^2$ for three different values of the Rabi energy and the
same optical detuning $\delta/t_0=0.05$ ($t_0=70$ meV) as a function
of the site number. We estimate that the value of the Rabi energy
needed for self-trapping corresponds to laser intensities of about 2.5
kW/cm$^2$, which is reasonable for nanosecond pulses. At the lowest
value of the Rabi energy the polarization is small and homogeneously
distributed along the wire. For $\Omega\sim 150 \mu eV$, the solution
is localized due to the phonon-assisted self-trapping.  When the Rabi
energy increases slightly, the polarization distribution becomes
broader and flatter at the top. This saturation is related to the
intrinsic fermionic nature of the excitons.  Notice that the total
polarization increases when the Rabi energy increases. A gradient in
the density of polarization produces a force on the ions according to
Eq.~(\ref{forceeq}). The force is stronger at edges of the
$|\psi_n|^2$ distribution as seen in Fig.~\ref{force}, and is positive
to the right from the center of the chain and negative to the
left. The quantity that is measured in electron or x-ray diffraction
experiments would be $\partial u_n/\partial n$, i.e. the local
deformation due to the light-induced changes in the lattice. For the
parameters considered above, this quantity is in the range 10-100
m\AA, which is reasonable for an experimental observation of the
effect.


We have neglected exciton-exciton Coulomb effects which may affect the
picture given here, especially in the high excitation regime.  This
deserves further investigations. However, it is known that the BCS
variational ansatz in Eq.~(\ref{var}) provides a good description of
the ground state of an electron-hole system both in the low excitation
regime (excitonic BEC) and in the high excitation regime (BCS
pairing).~\cite{comte82} The fact that the electron-hole system here
is driven by an external coherent field (laser) gives an additional
justification to this ansatz since light-dressing can reduce Coulomb
features beyond mean field like screening and
broadening.~\cite{ciuti00} Finally, exciton-exciton interaction in the
dynamic Stark shift can be neglected when the optical detuning is
bigger than the biexciton binding energy,~\cite{combescot88} therefore
its effect can always be reduced by increasing the optical detuning.

In conclusion, we have investigated the effect of an intense
off-resonant laser field in polydiacetylene chains. We have extended
the SSH model to include the effect of a control laser field
semiclassically. We have obtained a nonlinear inhomogeneous equation
describing the spatial distribution of the optical polarization in the
chain in the steady state. We find solutions describing self-trapping
of the polarization which saturate when the intensity of the field is
increased.  The results suggest that local steady state lattice
deformations can be finely controlled by the intensity and frequency
of an intense laser field. This control scheme could
have interesting applications to other organic materials where a
strong exciton-phonon interaction exists, like e.g. in J-aggregates, or
DNA.~\cite{conwell00}

This work has been supported by NSF under DMR-0312491. We thank
Prof. S. Billinge, Prof. C.Y. Ruan and Prof. J. Fern\'andez-Rossier
for stimulating discussions.

\end{document}